# Gaia as Solaris: An Alternative Default Evolutionary Trajectory

Srdja Janković
University Children's Hospital, Tiršova 10
Belgrade, RS

Ana Katić
University of Belgrade Faculty of Philosophy
Belgrade, RS

Milan M. Ćirković
Astronomical Observatory of Belgrade
Volgina 7
Belgrade, RS 11000
E-mail: mcirkovic@aob.rs ; phone: +381691687200

**Abstract**. Now that we know that Earth-like planets are ubiquitous in the universe, as well as that most of them are much older than the Earth, it is justified to ask to what extent evolutionary outcomes on other such planets are similar, or indeed commensurable, to the outcomes we perceive around us. In order to assess the degree of specialty or mediocrity of our trajectory of biospheric evolution, we need to take into account recent advances in theoretical astrobiology, in particular (i) establishing the history of habitable planets' formation in the Galaxy, and (ii) understanding the crucial importance of "Gaian" feedback loops and temporal windows for the interaction of early life with its physical environment. Hereby we consider an alternative macroevolutionary pathway that may result in tight functional integration of all sub-planetary ecosystems, eventually giving rise to a true superorganism at the biospheric level. The blueprint for a possible outcome of this scenario has been masterfully provided by the great Polish novelist Stanisław Lem in his 1961 novel *Solaris*. In fact, *Solaris* offers such a persuasive and powerful case for an "extremely strong" Gaia hypothesis that it is, arguably, high time to investigate it in a discursive astrobiological and philosophical context. In addition to novel predictions in the domain of potentially detectable biosignatures, some additional cognitive and heuristic benefits of studying such extreme cases of functional integration are briefly discussed.

The Context: Solaris, Macro-/Megaevolution, and the Taxonomy of Gaia Hypotheses

In nearly five decades that elapsed since the original Gaia hypothesis – claiming that the Earth's biosphere is a vast adaptive system capable of actively maintaining biogeochemical homeostasis, or indeed a kind of superorganism – was brought forward by James Lovelock and Lynn Margulis (Lovelock and Margulis, 1974), its relation to evolutionary theory has itself very much evolved and complexified. It has also been increasingly recognized that important problems in philosophical and scientific discussions of the Gaia hypothesis are posed by the fact that it is no longer a single hypothesis. As demonstrated by Doolittle, all meaningful attempts to prove a high or a low plausibility of Gaia must begin by making distinctions, at least, between "strong" and "weak" versions of Gaia, or a "probable" and "accidental" Gaia (Doolittle, 2017). These different versions arguably deserve different names or designations. If we are to attempt any such taxonomy, a good way to start might be to define the upper and lower limits of "Gaia-ness", *i.e.*, of biospheric/planetary integration level. While at the lower end Gaia blends with the simple recognition of ecological inter-relatedness of all organisms (and organisms with their environment), the *upper* end of the "Gaia spectrum" is much less explored. It might, therefore, be useful if, similar to Italo Calvino's Count of Monte Cristo, who tries to envisage the strongest conceivable prison in order to find a potential escape route from a real one (Calvino, [1967] 1969), we make an earnest attempt to imagine the strongest physically possible version of Gaia – that of some sort of superorganism endowed with consciousness and capability of intentional action. We will refer to this theoretically constructed extreme Gaia as Solaris, borrowing the term from Stanisław Lem and Andrei Arsenievich Tarkovsky (Lem, [1961] 2003; Tarkovsky, 1972).

Lem's Solaris is a planet orbiting two suns (in so-called *circumbinary* or P-type orbit, similar to Kepler-16b; Doyle *et al.*, 2011), which was originally detected via unusual *stability* of its orbit. The planet is almost entirely covered by global ocean, which manifests extremely complex behavior and is, among other things, the source of homeostatic influence on the planet's orbit. In other words, Solaris retains its stable orbit due to direct influences exerted by its global ocean. By most criteria, the ocean seems to be a living system, and – at least on some construals – an intelligent system. The whole scientific discipline of *solaristics* emerges in response to this epochal astrobiological discovery and is described in often quite painstaking detail in the novel.[1] These passages are a true goldmine for philosophy of science – especially philosophy of biology and cognitive science – but, as we shall try to show in this paper, for theoretical astrobiology as well. By the time the plot of the novel takes place, however, solaristics has largely degenerated into recording and categorizing the complex phenomena occurring at the surface of the ocean, without any unifying theoretical principles and with little hope of obtaining more general explanations.

The plot of the novel is relevant for us here only insofar it provides clues for the astrobiological background. A psychologist named Kris Kelvin arrives aboard the Solaris research station, hovering above the oceanic surface. Shortly before his arrival, the crew of the station conducted a more aggressive (and unauthorized) experiments with modulated X-ray bombardment in order to force the desired Contact. Their experimentation yields entirely unexpected results and becomes psychologically traumatic for themselves, as individually flawed humans. The ocean's response to the intrusion exposes deeper, hidden aspects of personalities of the *human* scientists, while revealing essentially nothing of itself, barring supreme capacity of manipulating matter at the level of elementary particles. To the extent that the ocean's actions can be understood (or indeed ascertained to be actions in the first place), it seems to test the minds of the scientists by confronting them with their most painful repressed memories. It does this through materialization of physical *simulacra*, including those resembling human beings; Kelvin confronts memories of his deceased wife and feelings of guilt about her suicide. There is no real resolution to the plot – in the same manner, it is tacitly suggested throughout the novel, as there is no resolution or goal to the cosmic evolution itself, just the eternal "turning and turning" of challenges and responses.

---

[1] Lem, [1961] (2003), esp. pp. 15-26, 115-130, and 171-183.

Now, what could Solaris be in the general astrobiological context of the real universe? For our present purposes, it is a complex biosphere – a Gaia – that has achieved the highest imaginable level of functional integration (with or without exploring further possibilities along "novel" dimensions in the parameter space, *e.g.*, advanced noögenesis, technological capacities, astroengineering, etc.). Elaborating possible evolutionary trajectories leading to a Solaris in a sufficiently general context, apart from its primary importance in determining the status of the Gaia hypothesis within the necessarily evolutionary framework of biological science, is almost certain to entail broad and deep epistemological insights. To tap into this potentially rich source of new understanding of processes, mechanisms, and constraints of biological evolution, we need to re-evaluate the key assumptions of relevant versions of the Gaia hypothesis given the emerging knowledge on macroevolutionary pathways, major evolutionary transitions, multilevel evolvability, and coding concepts that govern the flow (and enable the causal role) of information in biological systems. In the present paper, we will attempt to lay some groundwork for future efforts in this direction. This is in accordance with the important recent studies in astrobiology aiming (i) to make precise the notion of "Gaian feedbacks" necessary for evolving an Earth-type biosphere (Chopra and Lineweaver, 2016), and (ii) to affirm the "different is more" meta-principle in searching for other, not necessarily Earth-like habitats (Lenardic and Seales, 2019).

Life and its Code(s): The Little Apocrypha

As Carl Gustav Jung farsightedly commented in a 1928-1930 seminar, although living matter has no "special" properties in terms of its adherence to common laws of physics and chemistry, there must be some fundamental feature distinguishing the *processes* of a living body from all that is non-living (Jung, 1984). Almost a century later, armed with conceptual tools of information theory, we see this as a specific type of meta-pattern of information flow, ultimately stemming from the way information is encoded and processed – the *coding concept*. Thus the rapidly evolving field of code biology must be relevant to all considerations of biosphere-level processes, including functional integration (for an overview of code biology and a tentative taxonomy of various codes present in living systems, see Barbieri, 2018; for further reading about the information-based research approach see Hidalgo *et al.*, 2014). Another cornerstone of our analysis is the universally held, but not universally explicated assumption that frequencies of individual chemical reactions on early Earth (or its general equivalent) – including, but not limited to those of prebiotic relevance – follow(ed) the "large number of rare events" (LNRE) distribution (Hazen, 2017), since this type of distribution, given sufficient time, allows efficient "sampling" of conceptual/genetic/morphological/functional space, effectively providing natural selection with an operational substrate. While modern cosmology could help in this regard (Koonin, 2007; Totani 2020), especially with some additional support from the theory of observation-selection effects (Bostrom, 2002), it is not obvious that such help is indeed necessary. It may become so if it turns out that the universe is full of empty habitats: perfectly habitable planets which are not, however, inhabited (more on this below).

Thinking Gaia-wise, although there is no inherent mechanism to ensure that prebiotic/early biotic systems feature negative (homeostatic) rather than positive (destabilizing) feedback loops, natural selection will arguably tend to preserve the ones with strong homeostatic elements. Just as LEGO bricks have inherent capabilities that make possible increasingly complex arrangements (for a deeper discussion of information-related aspects of LEGO please see Crompton, 2012), so the last universal coding-conceptual ancestor (LUCoCoA) must have been endowed with certain capabilities making homeostatic regulation within the proto-biosphere of its provenience highly likely (or, at the very least, likely enough to offer reasonable odds of long-term survival). Prebiotic ensembles organized around coding concepts unable, or unlikely to be able, to give rise to homeostatic feedback regulation will, accordingly, be filtered out relatively early in planetary history. At the intersection of explanations of Gaia's multilevel evolution and information biology, our understanding of natural selection, in the light of contemporary evidence, involves non-standard evolutionary mechanisms such as variation/persistence models, and not exclusively population/reproductive models (more on this later).Arguably, the key segment of the prebiotic-to-biotic transition (and seemingly a midway point between chemical and biological evolution) is embodied in the formation of the primary biological code. This is referred to as "codepoiesis" by Barbieri and as "conception phase" by Janković and Ćirković (2016). This phase is also characterized by selection for potential future common ancestors: it is then that proto-life acquires the primary evolvability – the ability to open-endedly acquire abilities by subsequent evolutionary processes. Some of these abilities are virtually guaranteed to be relevant for the establishment of interaction networks, modular genome structures with further enhanced evolvability, biosphere-wide feedbacks, and ultimately, to our present interest, a (more or less) Gaian level of functional integration, potentially promoting long-term stability of planetary conditions and survivability. Along these lines, Chopra and Lineweaver (2016) have, in the aforementioned paper, already convincingly shown that the long-term habitability of a planet is critically dependent on what they called a "Gaian Window": an early establishment of planet-wide homeostatic mechanisms, enabling a certain degree of functional regulation of the proto-biosphere via biogeochemical feedback loops, such as those concerning temperature stabilization, atmospheric contents, cycling of key bioelements, etc. These authors argue that extinction soon after abiogenesis is the cosmic default, because life only rarely evolves quickly enough and manages to persist long enough through various biochemical aggregates/ensembles to provide the required feedbacks and thus allow the planet to remain habitable. In their model, therefore, our planet may well be atypical in its rapid evolution of biogeochemical regulation. (For full implications of this important feedback loop – the need for a planet to become inhabited in order to remain habitable – please see Zuluaga *et al.*, 2014.)

However, we are here considering the possibility that a significant Gaian bottleneck could lie in an even earlier era – for, having successfully negotiated the killing grounds of conception, the future LUCoCoA can be reasonably expected to already possess at least some of the key properties that form "the seed" of a prospective Gaia: at the time of first "organisms", the bottleneck might already have been negotiated, and life might already possess in-built capacities aiding it to endure, even though we do not have sufficient reason or evidence to claim this at present. Obviously, if the evolvability that had been built in the coding concept that eventually survived and entered the Chopra-Lineweaver bottleneck had any contribution to subsequent outcomes, it warrants exploration. The postulation of the conception phase also helps us to further operationalize the obvious conclusion that initiation of life (abiogenesis) and its evolution require a completely separate set of conditions. This would remove many confusions and obstacles in our understanding that can only aid and abet creationist screeds about "the Darwinian evolution not explaining abiogenesis" and the like. Translated into this conceptual framework, the currently dominant orientation of astrobiological research programs (*i. e.*, the idea that we should start from "life as we know it" and then potentially proceed to "life as we do not know it") could be (re)defined in terms of the section of conceptual parameter space corresponding to the Terran biosphere, as well as larger sections it is contained within (though these must not be viewed as concentric circles, for they are not necessarily circular *or* concentric, or indeed even contiguous – after all, most archipelagos do include small outlier islands).

These larger sections, of course, correspond to progressively broader generalizations of "our" life-concept, although at least some of them could actually branch off in a different direction within the n-dimensional parameter space of all possibilities (as embodied by the "library of libraries", where each "volume" corresponds to a complete Dennett's "Library of Mendel" covering a given coding concept; Janković and Ćirković, 2016).

In the arguably central chapter of *Solaris*, entitled "The Little Apocrypha", Lem iuxtaposes the present and the past in the most disconcerting, albeit instructive, manner (Lem, [1961] 2003, pp. 69-93). The chapter has two parts, sharply contrasted in both content and tone: the conversation of Kelvin and Snow, and Kelvin's reading of the "apocryphal" historical documents. The main protagonist, Kelvin, somewhat unwillingly engages in the key dialogue with his solarist colleague Snow, where the latter reveals his utter disenchantment with any prospects for human contact with an extraterrestrial intelligence. The conversation culminates in the famous anti-anthropocentric, drunken monologue of Snow: "We have no need of other worlds. We need mirrors." This is correctly regarded by literary critics and theorists (Peter Swirski and Zoran Živković among others) as the philosophical summit of the novel (Živković, 1983; Swirski, 2015). At the end of the same conversation, however, Kelvin obtains the eponymous obscure book which – as indicated by their deceased colleague Gibarian – could shed some light on their predicament.

Now, while "The Little Apocrypha" is the title of the book-within-book which contains vital *past* records – in this particular case, it is in fact the *present* which retroactively illuminates and explains the significance of Berton's report and other documents contained therein. (In Tarkovsky's movie, this is underscored by using a black-and-white video footage of Berton's testimony before the puzzled commission, which to the viewer occurs entirely inexplicable without *subsequent* information and Kelvin's experiences on Solaris.[2]) Therefore, the entire chapter points to the necessity of *double* emancipation: both from our default anthropocentrism in approaches to astrobiology/SETI studies *and* from a kind of historicism that blithely expects to find neat causal chains leading from the past evolutionary influences to the presently observable state of a biosphere. (For Lem's rather complex views on historicism, please see Ćirković, 2021).

Do (Almost) All Roads Lead to (Some Sort of) Gaia?

Having established at least the plausibility of evolutionary trajectories leading to some types of Gaia (including Solaris), it is time to consider their relative likelihood and potential universality in a cosmic context. In order to do this, we need to evaluate key steps necessary for the evolutionary emergence of a highly integrated biosphere, as recently summarized by Bains and Schulze-Makuch (2016) from the perspective of our origins. Knowing that the evolution of life on Earth proceeds through a series of evolutionary transitions, powered by the attainment of successive levels of self-organizing complexity, the question of early vs. late Gaian bottleneck can be reformulated as follows: what is the earliest level of biological organization that enables a critical degree of global homeostasis, *i.e.*, a Gaia-like quality. Since this is, arguably, a prerequisite for long-term survival (and therefore for the possibility of all subsequent, higher-level transitions), early levels certainly appear to be better candidates than late ones. Admittedly, this type of inference also has clear limitations, particularly while trying to generalize from a sample comprised of a single biosphere.

The most important counterargument to the present scenario is the one invoked by evolutionists, starting with Dawkins, against any strong version of the Gaia hypothesis: there are *a priori* no mechanisms to prevent "secession" or free-riding of lower-level component lineages contributing to Gaia, and this exerts definite fitness costs at those lower levels. This is a strong argument, but should not be overrated. The analysis put forward by Doolittle (2014) applies here as well: we can speak about selection at the biospheric level through extinction alone, so it should be possible to argue that Gaias without defectors are more successful in avoiding extinction than those with defectors. Furthermore, we may doubt the temporal scales involved in the process of defection, and doubly so: first, there is no obvious obstacle for defection to require, at least in some cases, a very long time, comparable to the cosmological timescales; second, it seems reasonable to assume that, similar to other cases of Gaian feedbacks as discussed by Chopra and Lineweaver (2016), there is only a finite temporal window in which defection is possible. After all, a similar defection argument could be applied to endosymbiosis: in a cartoonishly simplified way, one may ask why not even extreme heterodox views have ever argued that Bacteria and Archaea evolved, presumably by secession, within the clade of Eukarya (cf. Mariscal and Doolittle, 2015)? As emphasized by a recent introduction to a collection of papers on the topic (Keeling, McCutcheon, and Doolittle, 2015, p. 10101):

> The shift from a transient inhabitant of another cell to a permanent and essential organelle was likely a protracted series of events, moving through levels of integration with no particular endpoint. Somewhere along this spectrum there may have been a turning point, or a no-going-back moment, perhaps marked by a sudden change in the tempo or mode of evolution. Perhaps this key change happened so long ago that any unambiguous evidence has been too obscured by time. Or, perhaps there was no such moment.

---

[2] Solaris, [1972] (2011). See also Bold (2014) for a perceptive analysis of this scene.

It is precisely the consequences of the last option which we investigate here. That said, one should notice that the "secession" argument is perhaps too much grounded in our parochial, terrestrial experience to be of universal validity. Not only could we regard natural selection in Doolittle's manner as selection among Gaias in the Milky Way, but if interstellar panspermia occurs (for some of the recently rekindled interest in the topic, see Ginsburg, Lingam, and Loeb, 2018; Sadlok, 2020; Steele *et al*., 2019), there is no obstacle for having a full-fledged differential reproduction + extinction selection within the Galactic ecosystem.

Now, if we regard Gaia – as many do – simply as the ultimate level of symbiosis in the natural world, it is reasonable to expect that we may gain useful insights by studying the processes that generate and stabilize phylosymbiotic relationships across the tree of life on Earth. These processes include both coevolution *sensu stricto* and purely ecological filtering processes, as recently analyzed by Mazel and coworkers (2018) using computer simulations. Notably, their set of simulations demonstrated that examples of phylosymbiosis in the majority of taxa could likely have arisen by host-related ecological filtering. This result readily yields itself to the interpretation favorable to all versions of the Gaia hypothesis: namely, that the tendency to generate a biosphere imbued with a network of symbiotic relationships must be deeply embedded in the core biological processes, allowing at least the possibility to generalize outside our current example of life. Another important idea is contained in a recent study of niche emergence by Gatti *et al*. (2018). By representing biodiversity as inherently autocatalytic on every level – as a subset or as a whole – these authors go a step further towards the understanding of evolutionary processes as fundamentally dynamical and able to generate non-linear self-sustainability. The main difference between the well-accepted concept of complex adaptive system (CAS) and a new ecological concept of the autocatalytic set is in the interpretation of its capability to respond to the external perturbations, forcings, or drivers. While CAS refers to a passive reaction to exogenous impacts, autocatalysis indicates an essentially more active and dynamical role of every ecological component, individual or group, where "new niches can emerge as a result of species occupation following a fractal hypervolume expansion" (Gatti *et al*., 2018, p. 113). So, interactions among system components, at a local level, enable pattern formation that profoundly influences the system's further development. Thus we have (in space) autocatalytic ecological sets expanding (in time) and becoming autocatalytic evolutionary sets that are an integration of ecosystems. Another stride in a similar direction is the even stronger concept of *autopoiesis* (Rubin *et al.*, 2021).

Although the purpose of this paper is not to analyze how life came to being, but rather to increase the range of possible evolutionary pathways resulting in viable biospheres by including superorganism-strong functional integration at the biospheric level, we also find it useful to mention Bruce Damer's work in modeling the terrestrial origin of life hypotheses (Damer, 2016; Van Kranendonk *et al.*, 2017). Regardless of whether or not they will manage to experimentally and unequivocally prove their thesis soon, the one thing emerges as an unavoidable research conclusion: irrespectively of the generic probability of abiogenesis, when the first form of protobiotic microcommunity had come to life, it came in the *specific environmental context* and – crucially – as *collective*, *communicable*, and at least *in part random* endeavor. The emergence of life should be associated with this first microcommunity representing our common ancestor, rather than the first alive entity/thing, because it seems that it takes far less energy and sophistication to maintain a cooperative protocellular community than to ensure the survival of a clearly segregated individual in its challenging environment. In other words, it is the Black Queen hypothesis (Morris, Lenski, and Zinser, 2012) – the loss of genes by natural selection due to redundancy in co-evolving communities – rather than the Red Queen default, which offers incentives for tighter functional integration.

Finally, there might be a wider unifying philosophical principle in need of fleshing out here. In this unfolding epoch of multidisciplinary investigations, potential environmental catastrophic risks, and ever-evolving ability to comprehend the increasingly complex world around us (from micro-, across macro-, to mega-levels), we could remember Kant's idea of difference between constitutive and regulative principles (Kant, [1790] 1987). In this philosophical sense, Solaris could be not only the logical endpoint of meaningful Gaian hypotheses, but also the new regulative principle for us, as the farthest point we reach in considering the eco-evo theoretical synthesis. Unlike the Kantian idea of God, which, as an unchangeable, static, rooted principle, represents the horizon of our knowledge, Solaris is perhaps a dynamical principle that could evolve into the constitutive principle of our practical knowledge and our collective responsibilities. The relationship of what is "a fluke" vs. what is "the norm" becomes quite complex in light of modern cosmology, even before going into evolutionary complications (Olum, 2004; Koonin, 2007; Totani, 2020); adopting a wider, cosmological perspective, via astrobiology, in a wide range of human thought and practices could foster wise long-termism and have other beneficial consequences. We shall return to this theme in the concluding section.

Functional integration

If we take Lem's Solaris as the prototype of the highest level of functional integration of a global (planet-wide) biogeochemical system, then how is this highest level of integration to be attained? Obviously, this extreme integration does not preclude the existence of a rich internal structure, including semi-autonomous (or even fully autonomous) substructures. As Lem himself stated in the novel:

> The decision to categorize the ocean as a metamorph was not an arbitrary one. Its undulating surface was capable of generating extremely diverse formations which resembled nothing ever seen on Earth, and the function of these sudden eruptions of plasmic 'creativity', whether adaptive, explorative or what, remained an enigma (Lem, [1961] 2003, p. 23).

Thus, we face the following question: If the coding concept is favorable to the emergence of a Solaris, how would this be manifested in the morphological space of evolution? The classic theories of endosymbiosis (Margulis, 1967; Margulis and Bermudes, 1985) postulate several discrete steps of functional integration leading to the observed complexity of eukaryotes on Earth. Obviously, if analogous evolutionary processes occur elsewhere, on different Earth-like planets, it would be geocentric (and rather preposterous) to assume that the exactly same discrete steps must occur everywhere. In fact, there is no general reason why the process should ever cease – that is, until it encompasses the entire biosphere, resulting in a Solaris! In the classification of Bains and Schulze-Makuch (2016), eukaryogenesis on Earth is highly likely to have emerged through a "many paths" type of process. The same line of reasoning may be applied to other major evolutionary transitions as well – and different paths could be taken by different *Gaias in spe*. Of course, we also need to keep in mind that evolution is not directional, much less *uni*directional, and that the eventual arrival of any system into a Gaia-like (or a Solaris-like) state should by no means be assumed to occur without having negotiated a meandering path full of slowdowns or pauses, more or less sharp turns, digressions of variable magnitude, Borgesian furtively curved quasilinear sections, diverse tangential developments, and even temporary reversals. Even the general trend toward greater complexity is no more than a trend, and does not preclude possible dominance of processes that promote simplicity in some taxons or functional domains at any given epoch (O'Malley *et al.* 2016; for an illuminating discussion of physical foundations of biological complexity, please see Wolf *et al.*, 2018). The relevant kind of mutualistic interactions is best described by appeal to the Black Queen hypothesis, applied to the scale of the entire biosphere: all convergently-evolving redundant innovations will be subsumed under the single, global toolkit.

More practically speaking, there clearly are some initial/early conditions that could promote functional integration far more than it was the case on early Earth, hence leading to a much more functionally integrated Gaian system than the one we observe around us. In particular, we may envision two conditions being somewhat different:

- **Temporal margin:** functional integration of symbionts into the eukaryotic cell took very long time on Earth (at least 1.6 billion years, and possibly more; cf. Retallack *et al.*, 2013) and it occurred in ecological conditions very different from the ones characterizing abiogenesis. It is unclear at present whether this is a contingent or a convergent feature of our kind of Gaia. Since endosymbiosis and subsequent diversification of eukaryotes offer such a huge boost in fitness, it is at least reasonable to speculate whether it could happen much earlier on other worlds. Earlier occurrence would imply stronger functional links and more integrated Gaia (*i.e.*, a Solaris). Environmental factors built in the initial/boundary conditions such as the chemical composition of the protoplanetary material, the spectral type of the host star, the mass of the planet at the end of the accretion phase, the rate of impacting planetesimals, etc., could impact the duration of the relevant "Gaian window" in the Chopra and Lineweaver sense. Alternatively, for very long timescales (e.g., $10^{12}$ yrs or more), such as those characterizing hypothetical habitable planets around M-dwarf stars, the course of chemical, biochemical, and early biotic evolution may strongly diverge from the Earth blueprint, giving rise to a very different outcome, not (yet) observable anywhere at present, but perhaps dominant *sub specie aeternitatis*.
- **Topological structure**: whether functional integration will cease or continue throughout an ecosystem obviously depends on the spatial topology of that ecosystem. Suppose, for the sake of argument, that all life everywhere arises in Darwin's "warm little ponds" (cf. Damer, 2016). Since each pond is a little pond, there are obvious physical limits to the amount of matter which could be incorporated in living beings and the spread of symbiotic relationships between the emergent lifeforms. Here the topology of such ponds begins to play a role. Even if biochemistry stays exactly the same, unconnected and distant ponds would result in more diversity and less integration of early life than a connected (perhaps tidally influenced) system of many ponds in close proximity. Of course, this is just a fable: one should substitute whatever locale is the best candidate for abiogenesis from our best current theories, be it hydrothermal vents, radioactive beaches, clay deposits, etc. The general point remains, however: it is reasonable to expect that a more connected spatial topology of the region in which abiogenesis and early evolution of life proceed leads to stronger and more widespread functional integration.

Finally, while the institution of global homeostatic mechanisms undoubtedly requires cooperative relationships among lifeforms, it is often overlooked that this goes with a very broad definition of cooperation, one that is also bound to include competition and strife, as long as the overall result promotes and preserves long-term stability. In the immortal words of T. S. Eliot:

We [...] see the boarhound and the boar
pursue their pattern as before
but reconciled among the stars.

Striving for a deeper understanding of functional integration that has occurred in the deep past (questions of the origin of life) and may occur in the distant future (strong Gaia alternative), Bouchard's non-standard definition of fitness is insightful and fruitful. Considering that we lack precise definitions and complete models to understand biological populations and complex adaptations exclusively using the reproduction of individuals, he approaches the problem of the survival of the fittest differently. His arguments for understanding lineages temporarily through models of variability that persists versus population that reproduces not only offer the potential evolutionary mechanisms of clones within species and non-reproductive species, but also provide a new exploratory tool for further explaining global evolutionary transitions at the biosphere level. By bringing environmental fitness factors to the forefront, Bouchard's distinction between population and ensembles is of essential importance and thus sheds light on our understanding of biogeochemical conditions in protobiotic contexts (as we cannot analyze this topic in detail here, please see Bouchard 2011, 2014).

Solaristics: Further Evolution of a Solaris

Should one, then, regard Solaris as the proverbial "End of [evolutionary] history?" What could be said about the subsequent evolution of a Solaris? The eponymous Lem's case has been founded based on exactly what is a reasonably conceived abiotic selection pressure, namely the instability of planetary orbit and the need for its continuous correction by teleological or quasi-teleological action (Lem, [1961] 2003, pp. 15-16). This could be interpreted as the "highest level" of natural selection in a (Galactic?) population of biospheres, in the sense of Doolittle (2014) and Janković and Ćirković (2016). Entering the maximally integrative evolutionary "channel" leading to a Solaris would represent an extreme case of the Court Jester hypothesis (*e.g.*, Benton, 2009), since only the external, abiotic pressure could, under this assumption, account for the totality of morphological change. One can speculate that it is exactly the capacity for modifying its physical environment that acts to provide tighter integration within a superorganism of the Solaris type. There is no crucial qualitative difference between being capable of influencing planetary orbit, as described by Lem, and being able to detoxify harmful chemicals in a pool of stagnant water by an insect colony, in the well-known example discussed in detail by Wilson and Sober (1989).

While at this point we have a very poor basis to speculate about potential evolutionary paths leading beyond the described Solaris-state (in one or more meta-directions within some appropriately defined meta-space), it is wise to remember that we neither have a basis to assume beforehand that a Solaris would necessarily be the terminus of all evolutionary trajectories leading there; indeed, if our broad hypothesis that secular increase of biospheric evolvability promotes the survival and growth of proto-Gaias is correct, we need to also consider the issue in the opposite direction: should we not expect the strongest resulting Gaias to exhibit the highest overall levels of evolvability – allowing them, perhaps, even further transition(s) to who-knows-where? After all, Lem carefully left open the possibility that his "living ocean" did, in a sense, evolve a bit further through humans' attempt at contact, however unsuccessful by "our" standards. Be that as it may, a fully-fledged Solaris is not merely the highest imaginable or conceivable level of integration of a biosphere/Gaia, for it also entails the "noöspheric" element, both in the historical sense of Vladimir Ivanovich Vernadsky ([1926] 1986) and in the light of modern views on the evolution of intelligence and human society, as well as the potential postbiological evolution (Bostrom, 2014). This, in turn, may or may not require a distinct – even if partly overlapping – set of selection pressures that need to be elaborated separately. The point, to some extent, clearly applies to all "major transitions" in evolution (Maynard Smith and Szathmary, 1997). Notably, however, as argued by Fleming and Brandon (2015), a major transition may not be explicable by selection alone, for it may require a non-adaptive first step (a type of exaptation) in order to form the initial higher-level entities selection can then act upon. This exaptive first step toward a major transition (or, more technically, its principal prerequisite, *fitness discontinuity*) may be plausibly explained by the *zero-force evolutionary law* (McShea and Brandon, 2010; McShea *et al*., 2019). This, in turn, strengthens the explanatory role of evolvability, since evolvability is, by definition, positively correlated with the general propensity to exapt.

All this may offer a broader conceptual framework for current analyses and some of the debates regarding Anthropocene (and all the harm that human species is now inflicting on the biosphere), with obvious relevance to future studies. In particular, should it turn out that the same naturalistic paradigm can go a long way towards a successful explanation of such seemingly diverse issues as abiogenesis, evolution of intelligence (or consciousness), ecological integration/dynamical equilibria, interrelations between biological, cultural and technological evolution, as well as the (aggregate) anthropogenic threat to Earth's current ecosystems (with the consequent existential threat to humanity), such a paradigm will most certainly be a fruitful guidance in our current and future research programmes and directions of thinking. In even stronger terms, even if our kind of life/intelligence turns out to be a fluke on the cosmological and astrobiological scene, we could still reap invaluable benefits from insights into other, more common modes of being and evolving.

Discussion

Following the famous Confucian dictum about receiving a meal of fish vs. being taught how to catch fish oneself, while survival of any single external challenge is highly contingent, survival of multiple challenges, particularly if they are of disparate nature, will most likely be a result of the existence of general adaptive mechanisms, or, more importantly, of a *significant capability of their acquisition*. Likewise, the principal line of thinking in current attempts to "reconcile" Gaia hypothesis (at least in its weaker forms) with evolutionary principles is to try and demonstrate that biological systems prone to destabilize their environment will tend to be short-lived due to one or more types of selection effects. However, a fully satisfactory explanation will most likely have to include some complementary process(es) – namely, it may also be the case that ecosystems prone to evolve disruptive species or groups tend to last much less than those less disposed to evolve such agents and/or more strongly endowed (based on their general evolvability coupled with specific evolvability along the relevant evolutionary pathways) with the ability to evolve some sort of regulatory oversight at higher levels.

One of the strongest historical criticisms levelled at the Gaia hypothesis was based on presumable absence of competition between Gaia-level entities. As Richard Dawkins famously argued, for the Gaia hypothesis to work, we would have to see a "Universe full of dead planets" (Dawkins, 1982). At the time, this was received as a very strong argument against the possibility of Gaia arising by (classically interpreted) Darwinian evolution. However, in the light of the ongoing astrobiological revolution, notably the fact that we already know of thousands of extrasolar planets, with discoveries on an almost daily basis, this begs the question to what extent the "dead Universe" of Dawkins *was* so counterfactual. Recently, the possibility of "empty habitats" has been given renewed attention (*e.g.*, Cockell, 2011); however, the very concept is of necessity subject to observation-selection effects. What seems "empty" to us, may not be so in reality, and that pertains to biology no less than to physics (the notion of vacuum in quantum field theory being a pertinent example). Consider, for instance, the above-mentioned possibility of very slow abiogenesis on a hypothetical habitable planet around an extremely long-lived M-dwarf star. At the present early – in the cosmological sense! – epoch, such planet would appear exactly as a Dawkinsian "dead planet" or Cockell's empty habitat. And yet, in the truly inclusive astrobiological sense, such habitats, as Gaias-under-construction, could account for most of the total life in the universe *sub specie aeternitatis*.

Radically different Gaias could, therefore, escape our attention and pass as empty habitats, especially when observed across interstellar distances. The answer will, however, probably have to await our capabilities to probe the status of a significant sample of discovered planets or systems in terms of both the fulfillment of preconditions for life and the presence or absence of life itself (whether extant or extinct). More importantly, as Doolittle (2014) demonstrated, biotic competition between biospheres is *not* a necessary precondition for evolutionary mechanisms/trajectories leading to a Gaia (or a Solaris), for the same causal role can belong to incremental alterations in the survivability of the whole system over (cosmic) time, with filtering function readily provided by many sequential perturbations that the geobiochemical system as a whole is bound to sustain, including those dictated by various factors of its cosmic environment. To reiterate a metaphor featured in our previous work (Janković and Ćirković, 2016), it is almost as though Life incessantly reaffirms itself by playing a game of chess against Death (bringing to mind the archetypal figure of Ingmar Bergman's Antonius Block). And, like chess players, all biospheres (or life-concepts) will not show the same aptitude. Could it be that the early establishment of a proto-Gaia is one of the most forceful openings – or even *the setting of the chessboard* – in the evolutionary game for biospheres?

A useful approach to a broad and preliminary assessment of the level of generality (or generalizability) of Earth's particular example within the "Gaia-space" (a parameter space composed of all possible versions of Gaia, along the relevant theoretical axes; Figure 1) could be to ask the following question: how much of the available morphospace has Life on Earth already sampled? In other words, what is the relative size of the portion of the total evolvome based on the specific Terran coding concept that has already been "tried out", or turned into actuality? Attempts to answer this question are already underway (please see Dryden *et al.*, 2008; McLeish, 2015). The full answer is not only relevant to any and all discussions of convergence vs. contingency but also to aid the understanding of the potential structuring of the Gaia-space in terms of both evolutionary alleys and attractors. Such an analysis might help us prepare for the next level of analysis, where the key question becomes:

What is the exact position of the Terran Gaia within the total Gaia-space, and in what way, exactly, was that position determined (or subdetermined)?[3]

As proposed in the aforementioned paper by Doolittle (2017), a proper assessment of likelihood of establishment of biosphere-wide (or planet-wide) homeostatic mechanisms requires an analysis of global processes of natural selection acting upon collective ecosystem properties ("the song") instead of organisms or taxons ("the singers"). A detailed analysis in this sense was recently published by Lenton and coworkers (2018). Interestingly, Doolittle's "song not singer" approach took on a quite literal note in the recent modeling analysis of the global macroevolution of passerine song patterns (Pearse *et al.*, 2018). To no great surprise, these models have confirmed that broad-scale biogeographical factors played a very important role in birdsong evolution (within the phase space volume delimited by anatomical/morphological constraints). Notably, the tempo of song evolution in passerines had previously been linked to variation in developmental mode among lineages (Mason *et al.*, 2017). This is but an example of a speedily growing corpus of empirical data supporting a sort of paradigm shift in our understanding of macroevolutionary trends – a move towards a further extension of the "extended modern synthesis" to include, in addition to complex relationships between evolutionary and developmental processes (evo-devo) and ecosystem-related factors that codetermine evolutionary pathways (eco-evo), an explicit analysis of evolvability on multiple levels, including feedback in the form of "evolution of evolvability" (evo-evo or evo$^2$). It is important to note, however, that our analysis is largely independent of the exact construal(s) as to how life began, for the conception (or codepoietic) phase is where all hypothetical early prebiotic trajectories join. Even if different in actual processes, it will – conceptually – be a required step whatever the (physical) explanation for the origin of life (or the pre-conceptional prebiotic *ensemble*) may be. Conception would equally, in principle, be a required step if some elements of Panspermia hypothesis should happen to be correct – *i.e.*, some elements of the makeup of the prebiotic system entering the conception phase are of non-local origin. Broadly speaking, this requirement is already satisfied, even for the strictest propositions of local abiogenesis, given that Earth is an open system, and early Earth, with its absence of any life-driven homeostatic regulation, is certain to have been even more sensitive to any external disturbances; the conceptual basis of this line of reasoning can be readily generalized to any locale satisfying the broad conceivability criteria.

Another obvious and potentially strong point of criticism of what has been said so far rests on the complex relationship between genotype and phenotype, embodied by various types of phenotypic plasticity. While this is an extremely important direction of further inquiry (and one that cannot be disregarded if we strive toward a comprehensive explanation of macroevolutionary processes), it still holds that the relationship between genotype and phenotype (and thus between the concept and its realizations) must at all times be less than random. The many sources of non-linearity in this relationship are rather a vast reservoir of macroevolutionary variation and open-ended novelty, including the possibility to access "higher-dimensional" spaces that may also set the stage for some further key transitions. The same applies to horizontal gene transfer (HGT). In fact, HGT, or some generalization thereof outside of "our" coding concept, may be a prerequisite for successfully clearing the proto-Gaian temporal window, or at the very least, for a reasonably high (reasonably distant from zero) probability of such an outcome – since the temporal requirements of recruiting all the essential guilds of organisms by their independent first-order evolution (without the boost provided by the possibility of gene exchange and consequent quick co-option of taxonomically different groups) may indeed exceed the given physical limitations in most, if not all, early biotic locales (candidate biospheres/Gaias). In addition, there is no fundamental reason to *a priori* exclude the possibility of multiple, conceptually different organisms clearing the bottleneck in a conceivable locale and giving rise to multiple partial biospheres, *i. e.*, a composite biosphere/Gaia (or cohabitation of biospheres/Gaias). However, given that HGT played a critical role in stabilizing the homeostatic systems in the case of Earth, it is reasonable to expect that the power of integration (and thus the long-term survival potential) should be significantly greater within a single, all-encompassing (proto) Gaia.

---

[3] This position must be viewed longitudinally, *i.e.*, integrated over total time since conception.

An interesting twist, though not a necessary consequence of the described "pre-Gaian bottleneck", or "early Gaian bottleneck", might be a reversal of our current assessment of our own status within the biospheric evolution. Rather than as an evolutionary success, an intelligent, tool-making, noösphere-inaugurating species could be seen as a sort of failure of long-term Gaian homeostasis (since humans are the first species endowed with the capability to cause a mass extinction, or to severely damage its home biosphere, if not to outright destroy it[4]). This is in accordance with the argument, first put forward by George G. Simpson, that the intelligence of the human type might indeed have a huge *negative* adaptive value in the long run (Simpson, 1964). Simpson invokes the spectre of global nuclear war – and we could add a number of anthropogenic existential risks derived from the misuse of technologies unknown in Simpson's time (*e.g.*, Bostrom and Ćirković, 2008; Bostrom, 2013). Notice, however, that it is not intelligence *per se* which is problematic in these scenarios: it is intelligence coupled with other traits characterizing human species, traits like aggression, territoriality, or tribal preferences, which are clearly contingent upon the particular evolutionary pathway which led to the emergence of our species (Tinbergen, 1968). One could go even further to speculate that other pathways to intelligence and technological civilization (*e.g.*, Schwartzman and Middendorf, 2011) would be associated with far less risk of self-destruction, due to, for instance, higher innate aptitude in resolving coordination problems. A rather weak form of ergodicity would then lead to a predominance of such Gaias in the overall set. Quite plausibly, extremely integrated Gaias, such as Solaris, would be the most "skillful" in resolving the aforementioned challenges.

In a sort of counterpoint to the classical view that noösphere is a natural (and/or typical) extension of the biosphere[5] (highly relevant for post-biological evolution), we may just try out the contrary premise: that maintenance of strong integration, with prompt filtering out of all disruptive species, such as *Homo sapiens*, is the rule, and the development that took place on Earth is, then, an exception in this regard. (This is, of course, somewhat un-Copernican, but it is a good thing to remember that a typical entity will typically be *untypical* in a few of its many properties – since having *all* properties very close to the population median is actually not the most probable outcome; the section of probability space belonging to this configuration is rather narrow and flanked on both sides by much broader stripes of probability space occupied by entities typical in most, but untypical in one, two, three, etc. properties.) Therefore, perhaps unexpectedly, we may be faced with important consequences for the futures studies. High-level functional integration all over our planet has currently been erected by humans in various forms of economic, cultural, demographic, and other kinds of globalization. While it may not be self-sustainable in the sense of Solaris, important trends are leading in that direction. In particular, the Internet-of-Things indeed looks like an early stage of a wholesome, self-sustainable noösphere, where all information-processing functions, including those usually ascribed to psychology, would be integrated (Swan, 2012, 2013). This and other transhumanist tendencies have been envisioned by biologists (*e.g.*, Sagan and Margulis, 1987) and would be following the modern image of a "porous" barrier between biological and cultural evolution (Wagner and Rosen, 2014). It is at least conceivable that the ultimate endpoint of *both* types of evolution is a similar state of extreme functional integration, somewhat reminiscent of Teilhard de Chardin's "Omega point", whether it be achieved primarily through biological or through cultural (*i.e.*, technological) means, as in Bostrom's (2014) concept of a *singleton*. Interestingly, Lovelock himself appears to be thinking along these lines as well (Lovelock and Appleyard, 2019).

---

[4] This remains extremely unlikely given the resilience of many clades of microbes, as well as some macroscopic organisms, *e. g.*, tardigrades. In particular, deep lithospheric extremophiles living at km-order depths in the crust of our planet may be capable of surviving even the end of Solar evolution and complete congelation of Earth's surface (*e.g.*, Laughlin and Adams, 2000; McMahon, O'Malley-James, and Parnell, 2013).

[5] The view the authors of this paper still firmly hold at this time, pending further evidence.

**Figure 1**. The "Gaia-space": a metaspace allowing the classification of various Gaia hypothesis along the relevant theoretical axes. Rough putative positions of some familiar concepts of Gaia are also shown.

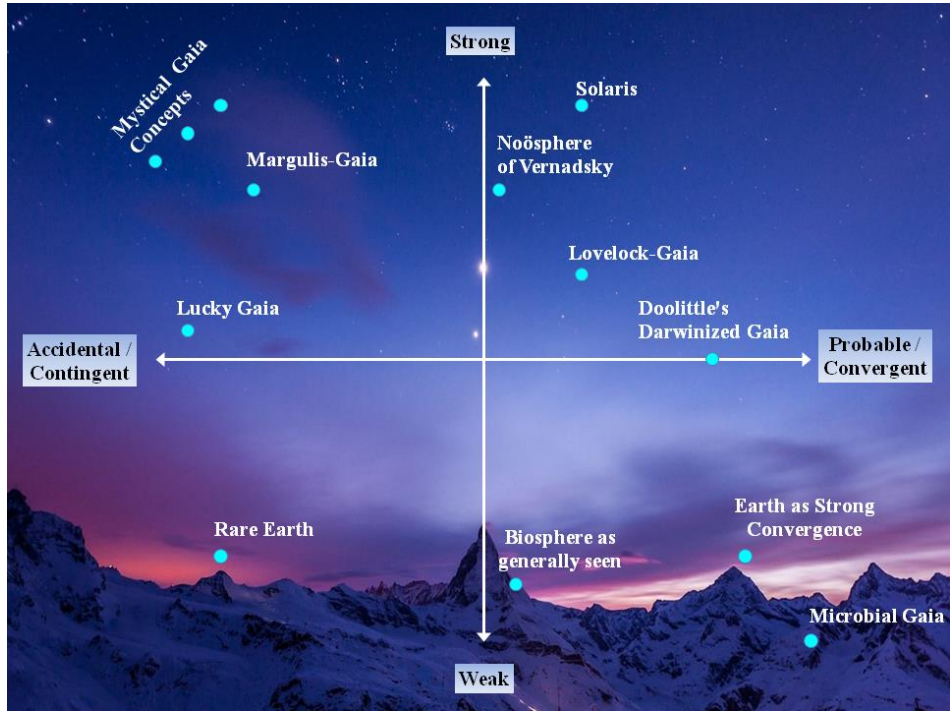

Conclusion

While all this could be seen as nothing but abstract and wild speculation, there is a number of quite practical consequences dealing with the current discussions of habitability (or conceivability) and the accompanying astrobiological search for habitable locales beyond Earth. Notably:

- We need to expand our searches for habitable locales with those cases where conventional wisdom would suggest marginal lack of prerequisites for habitability (in a sense, contrary to the "Rare Earth" requirements). Lem's idea about persistent response to environmental instabilities ought to be taken seriously in designing the searches for biomarkers.
- A class of models for prebiotic evolution with strong feedbacks and bottlenecks should be pursued somewhat beyond the usual region of parameter space representing the Earth-like domain of habitability.
- Functional integration, in particular, that is based on generalized (endo)symbiosis, should be modeled in a range of simulated paleoecological conditions, including very different timescales and multiple versions of the Black Queen hypothesis.
- Daisyworld-like models and numerical experiments should be investigated from the point of view of the conceivability criteria and the broad assumptions about Solaris-type Gaias. This pertains to a wider range of key parameters than hitherto assumed (*e.g.*, a wider range of allowed atmospheric chemical composition, surface morphologies, and physical parameters such as temperature), allowing for more robust Gaias and attempting to simulate potentially observable biosignatures. Any exploration along these lines would underscore the importance and value of *diversity* in habitability studies (as per precepts of e.g., Lenardic and Seales, 2019).
- If we assume that Gaia is the default direction in the general evolution of life and that Solaris is the default outcome in the evolution of a Gaia, this could also be a novel potential solution to Fermi's problem. (Although, admittedly, there would be considerable overlap with the Hermit and Persistence hypotheses; for classification, please see Webb, 2015; Ćirković, 2018.) This would also entail the necessity of adopting a new perspective on astrobiological complexity and its evolution (cf., Vukotić and Ćirković, 2012).
- The biotic-abiotic functional integration standpoint may be important for fully understanding the evolution of any planet, including a seemingly dead planet like Mars. This is particularly pertinent to the search for potential biosignatures, as exemplified by the recent analysis of Nathalie Cabrol (2018); after all, Lem's solaristics also began with biosignature analysis!
- If Solaris-type Gaias are numerically dominant in the set of inhabited planets, this would provide a neat resolution of Olum's (2004) anthropic problem: why do we find ourselves in a small civilization, when spatiotemporal considerations of cosmology suggest a typical observer to be located in a much older and bigger civilization? Well, if the largest fraction of civilizations consists of a single observer (or a single observer per planet), the problem is resolved.

In a scene that separates the prologue from the main part of Tarkovsky's film, the protagonist, Kelvin, acutely prepared to begin his journey towards Solaris, in a kind of bowiesque space-oddity moment asks the ground control: "Когда старт?" ["When do we launch?"]. He receives an almost immediate, calm response: "Уже летишь, Крис..." ["You are already on your way, Kris..."]. In a great work of art nothing exists by chance: we may well take this scene as a sharp reminder that, with the ongoing astrobiological revolution, we already are on our way to a new understanding of many "big questions" regarding Life in the Universe. And the associated challenge to our deeply-held views, inseparable from the the priceless opportunity to gain new perspectives, may indeed turn out to be more than a match for the complex experience of Lem's fictional (but all-too-real, and sometimes mirror-like) researchers in solaristics.

**Acknowledgements**. The authors wish to express warm gratitude to the two anonymous referees for OLEB for providing encouraging and thought-provoking comments on a previous version of this manuscript. M. M. Ć. wishes to thank Anders Sandberg, Zoran Živković, Milica Banović, Karl Schroeder, Bojan Stojanović, Mark Walker, Biljana Stojković, and Amedeo Balbi for pleasant discussions on related issues. A. K. would like to thank Vladan Ćurčić for inspiring criticism and insightful questions on this topic and Miroslav Đorđević for his wise intellectual support during the writing of this paper.